\documentclass[twocolumn,showpacs,preprintnumbers,amsmath,amssymb]{revtex4}

\usepackage{graphicx}
\usepackage{dcolumn}
\usepackage{bm}

\def\ii{{\mathrm{i}}}

\def\ee{{\mathrm{e}}}
\def\dd{{\mathrm{d}}}

\def\rot{\mathop{\mathrm{rot}}}

\def\sub#1{_{\mathrm{#1}}}
\def\up#1{^{\mathrm{#1}}}
\def\Vec#1{\mbox{\boldmath $#1$}}

\def\3He{\mbox{$^3$He}}
\def\4He{\mbox{$^4$He}}

\def\etal{{\it et al. }}

\begin{document}

\preprint{APS/123-QED}

\title{Quantum Turbulence in a Trapped Bose-Einstein Condensate}

\author{Michikazu Kobayashi}
\author{Makoto Tsubota}%
\affiliation{Department of Physics, Osaka City University, Sumiyoshi-Ku, Osaka 558-8585, Japan}%


\date{\today}

\begin{abstract}
We study quantum turbulence in trapped Bose-Einstein condensates by numerically solving the Gross-Pitaevskii equation.
Combining rotations around two axes, we successfully induce quantum turbulent state in which quantized vortices are not crystallized but tangled.
The obtained spectrum of the incompressible kinetic energy is consistent with the Kolmogorov law, the most important statistical law in turbulence.
\end{abstract}

\maketitle

The study of turbulence has a very long history, going back at least to Leonardo da Vinci, and understanding and controlling turbulence are great dreams of science and technology.
Classical turbulence (CT) exhibits highly complicated configurations of eddies.
Many studies have been devoted to the dynamical and statistical properties of CT after Kolmogorov's pioneering work \cite{Kolmogorov-1,Kolmogorov-2} on flow at very high Reynolds number, namely, fully developed CT.
The characteristic behavior of CT has been believed to be sustained by the Richardson cascade of eddies from large to small scales.
However, in CT, there is no universal way to identify each eddy, because they continue to nucleate, diffuse, and disappear.
As a result, many aspects of CT are still not perfectly understood.

Turbulence is also possible in superfluids, such as the superfluid phases of \4He and \3He.
Such quantum turbulence (QT) consists of definite topological defects known as quantized vortices and has recently attracted interest as a way to better understand turbulence \cite{Niemela}.

Superfluid \4He has been extensively studied, in particular with relation to quantized vortices \cite{Donnelly}.
Below the lambda temperature $T_{\lambda}=2.17$ K, liquid \4He enters the superfluid state through Bose-Einstein condensation.
The hydrodynamics of superfluid \4He is strongly influenced by quantum effects; any rotational motion is sustained by quantized vortices with quantized circulation $\kappa=\hbar/m$, where $m$ is the particle mass.
There are two typical cooperative phenomena of quantized vortices.
One is a vortex lattice under rotation in which straight quantized vortices form a triangular lattice along the rotation axis \cite{Yarmchuk}.
The other is a vortex tangle in QT in which vortices become tangled in a flow \cite{Vinen,Schwarz}.

QT has been studied as a problem in low temperature physics since its discovery some 50 years ago.
Its study has recently entered a new stage beyond low temperature physics.
One of the main motivations of recent studies is to investigate the relationship between QT and CT.
Some similarities between the two types of turbulence have been experimentally observed in superfluid \4He \cite{Maurer,Stalp} and \3He \cite{Finne,Bradley}, and have been theoretically confirmed by numerical simulations of the quantized vortex-filament model \cite{Araki} and a model using the Gross-Pitaevskii (GP) equation \cite{Nore,Kobayashi-1,Kobayashi-2,Parker}.
In particular, we have successfully obtained the Kolmogorov law for QT, which is one of the most important statistical laws in CT \cite{Frisch} by a numerical simulation of the GP equation \cite{Kobayashi-1,Kobayashi-2}.

The similarity between QT and CT means that QT is an ideal prototype to study the statistics and vortex dynamics of turbulence, because QT exhibits a real cascade process of quantized vortices.
However, in superfluid helium, it is very difficult to experimentally control the turbulent state and determine the vortex configuration.

Another important example of quantized vortices is magnetically or optically trapped atomic Bose-Einstein condensates (BECs) \cite{Pethick}.
The characteristics of trapped BECs are as follows: (i) a BEC system is weakly interacting and can be easily treated theoretically, (ii) many physical parameters of BECs are experimentally controllable, and (iii) various physical quantities such as the density and phase of BECs can be directly observed, which is in stark contrast to superfluid helium systems.
Quantized vortices can be considered to be holes of density and singularities of phase.
Shortly after trapped BECs were first realized, experimental groups \cite{Madison,Abo-Shaeer} reported vortex lattice structures, as well as the crystallization dynamics of these structures under rotation.
These dynamics have been successfully confirmed quantitatively by the numerical simulation of the GP equation \cite{Kasamatsu-2,Kasamatsu-3}.

However, in the experimental research of trapped BECs, another important phenomenon of quantized vortices, namely QT, has not been adequately studied.
Noting that quantized vortices are observable and that almost all physical parameters of trapped BECs are controllable, such systems are an ideal prototype for truly controllable QT, which is not possible for superfluid helium.
QT in trapped BECs can be used to determine several details of the system, such as the distribution of vortex length, details on the cascade of vortices, the isotropy or anisotropy of vortex configuration and details on correlations among vortices related to eddy viscosity, as already considered for CT \cite{Frisch}.
Clarifying any of these will allow the transition to QT to be considered a universal phase transition.
Therefore, research into QT offers the promise of greater advances in understanding turbulence than has been possible in past studies of turbulence.

There are only a few theoretical pioneering works for QT in trapped BECs in which QT is realized in relaxation processes from non-equilibrium initial state across the BEC critical temperature \cite{Svistunov}, or from vortex-free to vortex lattice state under rotation \cite{Parker}.
In this paper, we present a numerical simulation of the 3-dimensional GP equation.
First, we present the more experimentally realizable steady QT in a trapped BEC by combining rotations around two axes.
Second, we show that the spectrum of the incompressible kinetic energy $E\sub{kin}\up{i}(k)$ per unit mass obeys the Kolmogorov law
\begin{equation}
E\sub{kin}\up{i}(k)=C\varepsilon^{2/3}k^{-5/3}.\label{eq-Kolmogorov}
\end{equation}
Here, the energy spectrum is defined as $E\sub{kin}\up{i}=\int\dd k\:E\sub{kin}\up{i}(k)$, where $k$ is the wave number from the Fourier transformation of the incompressible velocity field and $\varepsilon$ is the energy transportation rate from small to large $k$ of the incompressible kinetic energy per unit mass.
The Kolmogorov constant $C$ is a dimensionless parameter of the order of unity in CT.
We further obtain the length distribution of vortices with a scaling structure that reflects the self-similarity in the Richardson cascade.

\begin{figure}[htb]
\centering
\includegraphics[width=0.5\linewidth]{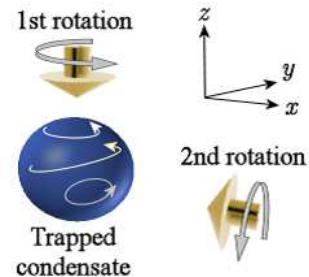}
\caption{\label{fig-rotating-BEC}(color online). Schematic sketch of the rotation. The first rotation is applied along the z-axis and the second rotation is applied along the x-axis.}
\end{figure}

In considering a trapped BEC system, we start from the GP equation
\begin{align}[\ii&-\gamma(\Vec{x})]\hbar\frac{\partial}{\partial t}\Phi(\Vec{x},t)=\Bigg[-\frac{\hbar^2}{2m}\nabla^2-\mu(t)\nonumber\\+&g|\Phi(\Vec{x},t)|^2+U(\Vec{x})-\Vec{\Omega}(t)\cdot\Vec{L}(\Vec{x})\Bigg]\Phi(\Vec{x},t).\label{eq-GP}\end{align}
Here $\Phi(\Vec{x},t)=f(\Vec{x},t)\ee^{\ii\phi(\Vec{x},t)}$ is the macroscopic wave function of the BEC, $m$ is the particle mass, $\mu$ is the chemical potential, $\Vec{L}(\Vec{x})=-\ii\hbar\Vec{x}\times\nabla$ is the angular momentum, and $g=4\pi\hbar^2a/m$ is the coupling constant with $s$-wave scattering length $a$.
The trapping potential $U(\Vec{x})$ is given by a weakly elliptical harmonic potential:
\begin{align}U(\Vec{x})=\frac{m\omega^2}{2}[(1-\delta_1)(1-\delta_2)x^2\nonumber\\+(1+\delta_1)(1-\delta_2)y^2+(1+\delta_2)z^2],\label{eq-harmonic-potential}\end{align}
where $\omega$ is the frequency of the harmonic trap and the parameters $\delta_1$ and $\delta_2$ exhibit elliptical deformation in the $xy$- and $zx$-planes.
To develop the BEC to a turbulent state rather than a vortex lattice state, we combine two rotations along the $z$-and $x$-axes, as shown in Fig. \ref{fig-rotating-BEC}.
The rotation vector $\Vec{\Omega}(t)$ is given by $\Vec{\Omega}(t)=(\Omega_x,\Omega_z\sin\Omega_xt,\Omega_z\cos\Omega_xt)$, where $\Omega_z$ and $\Omega_x$ are the frequencies of the first and second rotation, respectively.
There are two main advantages of using the combined rotation to study turbulence.
First, we can do direct numerical simulations and directly compare them with experiments without any ambiguity.
Second, it is possible to freely control the state from non-turbulent vortex lattice to fully-developed turbulence by changing the ratio $\Omega_x/\Omega_z$.
In the classical fluid system, Goto \etal have already adopted this combined rotation to study turbulence by using water in spinning sphere on a rotating turntable \cite{Goto} and reported the transition from rigid body rotation to non periodic turbulent motion of water.

The condensate density $\rho(\Vec{x},t)$ and the superfluid velocity $\Vec{v}(\Vec{x},t)$ are given by $\rho(\Vec{x},t)=f^2(\Vec{x},t)$ and $\Vec{v}(\Vec{x},t)=\hbar/m\nabla\phi(\Vec{x},t)$.
The vorticity $\rot\Vec{v}(\Vec{x},t)$ vanishes everywhere in a single-connected region of the fluid; any rotational flow is carried only by quantized vortices in the core, of which $\Phi(\Vec{x},t)$ vanishes so that the circulation is quantized by $\kappa=2\pi\hbar/m$.
The vortex core size is given by the healing length $\xi=\hbar/\sqrt{2mg\rho}$.
In trapped BECs, the healing length depends on the position, because the system is not uniform.
In this work, we define the characteristic healing length $\xi=\hbar/\sqrt{2mg\rho_0}$ with the condensate density $\rho_0=\rho(\Vec{x}=0)$ at the trap center.

We take a system at very low temperatures.
The phenomenological damping term $\gamma(\Vec{x})$ simulates the effect of thermal excitations and is effective only at scales smaller than the core size of quantized vortices, as shown in our previous work on the numerical simulation of a coupled system using the GP equation and the Bogoliubov-de Gennes equation (see Fig. 2(b) in the reference \cite{Kobayashi-4}).
In this work, we adopt the Fourier transformed damping term $\gamma(\Vec{k})=\gamma(k)$ at the temperature $T=0.01T\sub{c}$ given in our previous work, where $T\sub{c}$ is the BEC critical temperature of ideal Bose gas.
Introduction of $\gamma(\Vec{x})$ conserves neither the energy nor the number of particles.
To avoid the inconservation of the number of particles, we consider the time dependence of the chemical potential so that the total number $N=\int\dd\Vec{x}\:|\Phi(\Vec{x},t)|^2$ can be conserved.

To solve the GP equation (\ref{eq-GP}) numerically with high accuracy, we use the pseudo-spectral method, applying the Chebyshev tau method in space with a Dirichlet boundary condition in a box \cite{Boyd} containing $512^3$ grid points.
For the numerical parameters, we use the following, taken from experiments on $^{87}$Rb atoms \cite{Pethick,Madison}: $m=1.46\times 10^{-25}$ kg, $a=5.61$ nm, $N=2.50\times 10^{5}$, $\omega=150\times 2\pi$ Hz.
The total volume of the numerical box is set to $V=14.0^3$ $\mu$m$^3$.

We start from a stationary solution without rotation and elliptical deformation.
At $t=0$, we turn on the rotation $\Omega_x=\Omega_z=0.6$ and elliptical deformation $\delta_1=\delta_2=0.025$, and numerically calculate the time development of the GP equation (\ref{eq-GP}) using the 4th ordered Runge-Kutta method \cite{Boyd} with a time resolution of $\Delta t=1\times 10^{-4}\omega$.

\begin{figure}[htb]
\centering
\includegraphics[width=0.99\linewidth]{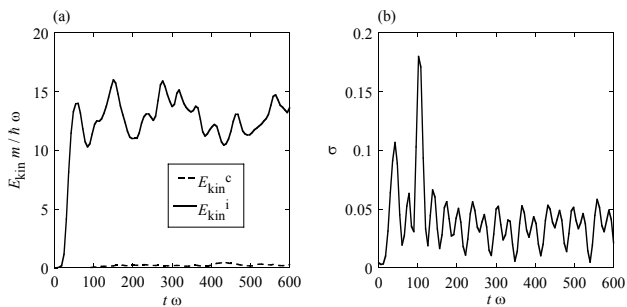}
\caption{\label{fig-sigma}Time development of $E\sub{kin}\up{c,i}(t)$ (a) and $\sigma(t)$ (b)}
\end{figure}

We calculate the total compressible and incompressible kinetic energy per unit mass $E\sub{kin}\up{c}(t)$ and $E\sub{kin}\up{i}(t)$ defined by
\begin{equation}
E\sub{kin}\up{c,i}(t)=\frac{1}{2N}\int\dd\Vec{x}\:\{[\Vec{p}(\Vec{x},t)]\up{c,i}\}^2.\label{eq-kinetic-energy}
\end{equation}
Here, $\Vec{p}(\Vec{x},t)=\hbar/mf(\Vec{x},t)\nabla\phi(\Vec{x},t)$, $[\cdots]\up{c}$ denotes the compressible part $\nabla\times[\cdots]\up{c}=0$ and $[\cdots]\up{i}$ denotes the incompressible part $\nabla\cdot[\cdots]\up{i}=0$.
We also investigate the anisotropy of the system by defining a parameter $\sigma(t)$ as
\begin{subequations}\label{eq-anisotropic-parameter}
\begin{align}
A(t)&=\int\dd\Vec{x}\:|\nabla\Phi(\Vec{x},t)|^2,\\
F(t)&=\sum_{i=1}^{3}\Bigg[\int\dd\Vec{x}\:|\frac{\partial}{\partial x_i}\Phi(\Vec{x},t)|^2-\frac{A(t)}{3}\Bigg]^2,\\
\sigma(t)&=\frac{F(t)}{A(t)},
\end{align}
\end{subequations}
for $x_{1,2,3}=\{x,y,z\}$.
Figures \ref{fig-sigma}(a) and (b) show the time development of $E\sub{kin}\up{c,i}(t)$ and $\sigma(t)$ respectively.
After $t\omega\simeq 150$, $\sigma(t)$ becomes small showing that the BEC recovers isotropy, and $E\sub{kin}\up{i}(t)$ becomes almost steady, which means that isotropic steady turbulence is realized at $t\omega\gtrsim 150$.
The steady turbulence is sustained by the balance between the large-scale energy injection due to the rotation and the small-scale dissipation.
The time of $t\omega=150$ corresponds to $t=0.1$ sec which is sufficiently shorter than the actual lifetime of trapped BECs in experiment.
$E\sub{kin}\up{i}(t)$ is always much larger than $E\sub{kin}\up{c}(t)$ and dynamics of the BEC is dominated by vortices rather than compressible excitations in the BEC.

\begin{figure}[htb]
\centering
\includegraphics[width=0.99\linewidth]{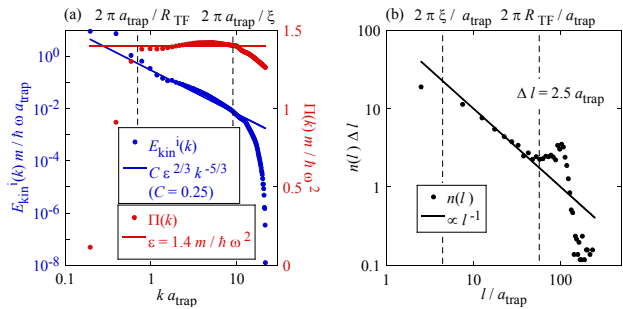}
\caption{\label{fig-energy-spectrum}(color online). (a) Wave number dependence of $E\sub{kin}\up{i}(k,t)$ and $\Pi(k,t)$. (b) Vortex Length distribution $n(l)\Delta l$ inside the Thomas-Fermi radius $R\sub{TF}$. $E\sub{kin}\up{i}(k,t)$, $\Pi(k,t)$ and $n(l)\Delta l$ are obtained from an ensemble average of 25 turbulent states ($t\omega >300$). In both figures, $a\sub{trap}=\sqrt{\hbar/m\omega}$ is the characteristic scale of the trap.}
\end{figure}

To confirm that the system is genuine turbulence, we calculate the spectrum $E\sub{kin}\up{i}(k,t)$ and the flux $\Pi(k,t)$ from small to large $k$ of the incompressible kinetic energy $E\sub{kin}\up{i}(t)$.
$\Pi(k,t)$ can be obtained by considering the scale-by-scale energy budget equation for the GP equation and given as (see Eq. (35) in the reference \cite{Kobayashi-2})
\begin{align} \Pi(k,t)=\frac{1}{N}\int\dd&\Vec{x}\:L_k[\{\Vec{p}(\Vec{x},t)\cdot\nabla\Vec{v}(\Vec{x},t)\}\up{i}]\nonumber\\ &\cdot L_k[\{\Vec{p}(\Vec{x},t)\}\up{i}].\label{eq-energy-flux} \end{align}
Here, $L_k$ is the operator for the low-pass filter:
\begin{equation}
L_k[s(\Vec{x})]=\frac{1}{V}\sum_{|\Vec{k}|<k}\int\dd\Vec{x}^\prime\:\ee^{\ii\Vec{k}\cdot(\Vec{x}-\Vec{x}^\prime)}s(\Vec{x}),\label{eq-low-pass-filter}
\end{equation}
for an arbitrary function $s(\Vec{x})$.
Figure \ref{fig-energy-spectrum} (a) shows $E\sub{kin}\up{i}(k,t)$ and $\Pi(k,t)$ for the turbulent state.
$E\sub{kin}\up{i}(k,t)$ in this QT satisfies the Kolmogorov law of Eq. (\ref{eq-Kolmogorov}) in the inertial range $2\pi/R\sub{TF}<k<2\pi/\xi$, where $R\sub{TF}=\sqrt{2\mu(t=0)/m\omega^2}$ is the Thomas-Fermi radius and represents the largest scale in the BEC \cite{BEC-expand}.
Furthermore, the energy flux is nearly constant value $\Pi(k,t)\simeq1.4\hbar\omega^2/m$ in the inertial range, supporting that the incompressible kinetic energy steadily flow in wave number space through the Richardson cascade at the constant energy transportation rate $\varepsilon=\Pi(k,t)$ in Eq. (\ref{eq-Kolmogorov}).
Using this $\varepsilon$, we obtain the Kolmogorov constant $C=0.25\pm0.2$ which is smaller than that in CT and consistent with our previous work for QT in the uniform system \cite{Kobayashi-1,Kobayashi-2}.

To investigate the relation between the Kolmogorov law and the Richardson cascade, we calculate the vortex length distribution $n(l)\Delta l$ inside the condensate (Fig. \ref{fig-energy-spectrum} (b)), where $n(l)\Delta l$ represents the number of vortices with length from $l$ to $l+\Delta l$.
At the turbulent state, $n(l)\Delta l$ obeys the scaling property $n(l)\Delta l\propto l^{-\alpha}$ for $2\pi\xi<l<2\pi R\sub{TF}$.
This reflects the self-similar Richardson cascade in which large vortices entering the condensate from the surface \cite{Kasamatsu-2,Kasamatsu-3} are divided into smaller vortices, which is first confirmed in turbulence with the framework of the Gross-Pitaevskii equation.
The scaling exponent $\alpha$ is close to unity, which is consistent with those given by Araki \etal ($\alpha\simeq 1.34$) \cite{Araki} and Mitani \etal ($\alpha\simeq 1$) \cite{Mitani}.

\begin{figure}
\centering
\includegraphics[width=0.9\linewidth]{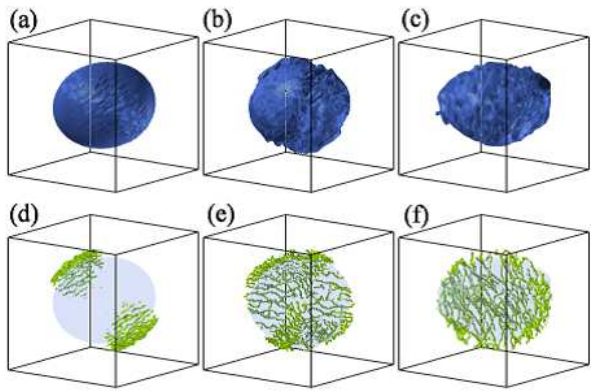}
\caption{\label{fig-turbulence}(color online). Isosurface plots of 5\% of the maximum condensate density (a)-(c) and configuration of quantized vortices inside the Thomas-Fermi radius $R\sub{TF}$ (d)-(f). (a), (d) $t\omega=10$; (b), (e) $t\omega=50$; (c), (f) $t\omega=300$. The method for identifying vortices in (d)-(f) is the same as that in Fig. 7 in the reference \cite{Kobayashi-2}.}
\end{figure}

To visualize the turbulence, we plot the isosurface of the condensate density $\rho(\Vec{x},t)$ and the spatial distribution of the vortices inside the condensate in Fig. \ref{fig-turbulence} (a)-(f).
At $t\omega=10$, the surface of the BEC becomes unstable (Fig. \ref{fig-turbulence} (a) and (d)), and vortices appear in the BEC at $t\omega=50$ (Fig. \ref{fig-turbulence} (b) and (e)).
Figures \ref{fig-turbulence} (c) and (f) shows QT with no crystallization but with highly tangled quantized vortices at $t\omega=300$.

In conclusion, using a numerical simulation of the GP equation, we have induced QT in a trapped BEC by combining rotations around two axes.
The quantized vortices in the trapped BEC are not crystallized but tangled.
In the inertial range, the spectrum of the incompressible kinetic energy obeys the Kolmogorov law and the energy flux becomes constant value.
We further obtained a vortex length distribution with a scaling property, supporting the self-similar Richardson cascade of quantized vortices.
The incompressible kinetic energy and its spectrum can be experimentally observed by measuring the density and phase of the BEC, according to Eq. (\ref{eq-kinetic-energy}).
We anticipate the experimental realisation of QT in a trapped BEC and further advancement of the understanding of turbulence.

The obtained energy spectrum in Fig. \ref{fig-energy-spectrum} (a) is not so destructive straight line and its consistency with the Kolmogorov law is incomplete.
This inconsistency comes from the anisotropy of turbulence around the y-axis around which there is no rotation, and is resolved by other simulations of quantum turbulence of trapped BECs under the combined rotations around {\it three} axes.
We will report on this study in near future.

MK acknowledges the support of a Grant-in-Aid for Young Scientists from JSPS (Grant No. 18840036).
MT acknowledges the support of a Grant-in Aid for Scientific Research from JSPS (Grant No. 18340109) and a Grant-in-Aid for Scientific Research on Priority Areas from MEXT (Grant No. 17071008).


\begin{thebibliography}{99}
\bibitem{BEC-expand}
The reason why $E\sub{kin}\up{i}(k)$ has large values for $k<2\pi/R\sub{TF}$ is that the condensate is expanded from $R\sub{TF}$ by the rotation.
\end{thebibliography}
\end{document}